\documentclass[aps,pre,reprint,footinbib,showpacs,superscriptaddress]{revtex4-1}
\usepackage{graphicx}
\usepackage[normalem]{ulem}
\usepackage{dcolumn}
\usepackage{bm}
\usepackage{amssymb}
\usepackage{color}
\usepackage{amsmath}
\usepackage{amsfonts}
\usepackage{soul}

\newcommand{\be}{\begin{equation}}
\newcommand{\ee}{\end{equation}}
\newcommand{\bea}{\begin{eqnarray}}
\newcommand{\eea}{\end{eqnarray}}
\newcommand{\bse}{\begin{subequations}}
\newcommand{\ese}{\end{subequations}}

\newcommand{\bx}{\mathbf{x}}
\newcommand{\bj}{\mathbf{j}}
\newcommand{\bJ}{\mathbf{J}_{tot}}

\renewcommand{\bm}{\mathbf{m}}

\renewcommand{\th}{\tilde h}
\newcommand{\comment}[1]{}

\begin{document}

\title{Universality classes for unstable crystal growth}

\author{Sofia Biagi}
\email{sofia.biagi@ujf-grenoble.fr}
\affiliation{Universit\'e Grenoble 1/CNRS, LIPhy UMR 5588, Grenoble, F-38401, France}
\affiliation{Istituto dei Sistemi Complessi, Consiglio Nazionale delle Ricerche, Via Madonna del Piano 10, 50019 Sesto Fiorentino, Italy}
\author{Chaouqi Misbah}
\email{chaouqi.misbah@ujf-grenoble.fr}
\affiliation{Universit\'e Grenoble 1/CNRS, LIPhy UMR 5588, Grenoble, F-38401, France}
\affiliation{Istituto dei Sistemi Complessi, Consiglio Nazionale delle Ricerche, Via Madonna del Piano 10, 50019 Sesto Fiorentino, Italy}
\author{Paolo Politi}
\email{paolo.politi@isc.cnr.it}
\affiliation{Istituto dei Sistemi Complessi, Consiglio Nazionale delle Ricerche, Via Madonna del Piano 10, 50019 Sesto Fiorentino, Italy}
\affiliation{INFN Sezione di Firenze, via G. Sansone 1, 50019 Sesto Fiorentino, Italy}

\date{\today}


\begin{abstract}
Universality has been a key concept for the classification of equilibrium critical phenomena, allowing 
associations among different physical processes and models. When dealing with non-equilibrium problems, however,
the distinction in universality classes is not as clear and few are the examples, as phase separation and kinetic roughening, for which universality has allowed to classify results in a general spirit. Here we focus
on an out-of-equilibrium case, unstable crystal growth, lying in between phase ordering and
pattern formation. We consider a well established 2+1 dimensional
family of continuum nonlinear equations for the local
height $h(\bx,t)$ of a crystal surface having the general form 
${\partial_t h(\mathbf{x},t)} = -\mathbf{\nabla}\cdot {[\mathbf{j}(\nabla h) +
\mathbf{\nabla}(\nabla^2 h)]}$: $\mathbf{j}(\nabla h)$ is an arbitrary function,
which is linear for small $\nabla h$, and whose structure expresses instabilities which lead to the formation of pyramid-like
structures of planar size $L$ and height $H$.
Our task is the choice and calculation of the quantities that can operate as critical exponents, together with the discussion
of what is relevant or not to the definition of our universality class.
These aims are achieved by means of a perturbative, multiscale analysis of our model, leading to phase diffusion equations
whose diffusion coefficients encapsulate all relevant informations on dynamics.
We identify two critical exponents:
i) the coarsening exponent, $n$, controlling
the increase in time of the typical size of the pattern, $L\sim t^n$; ii) the exponent $\beta$, controlling
the increase in time of the typical slope of the pattern, $M \sim t^{\beta}$ where $M\approx H/L$.
Our study reveals that there are only two different universality classes, according to the presence ($n=1/3$, $\beta=0$) or the absence ($n=1/4$, $\beta > 0$) of faceting. The symmetry of the pattern, as well as the symmetry of the surface mass current $\mathbf{j}(\nabla h)$ and its precise functional form, is irrelevant. Our analysis seems to support the idea that also space dimensionality is irrelevant.
\newline
\end{abstract}

\maketitle

\section{Introduction: universality classes}

The concept of universality is very useful in physics, because it
allows to classify seemingly different phenomena and models. 
Perhaps, one of the oldest examples is the
universal form of the Van der Waals equation of state
(law of corresponding states~\cite{Huang}), which is the simplest
equation describing a change of state and valid for any fluid.
A clear formalization of universality was firstly possible for equilibrium
critical phenomena, where the renormalization group theory allows to give
 a rigorous definition  of which parameters are relevant (universal) and which
are not. For example, within important classes of ferromagnetic spin
models, it is known that relevant parameters are:
the physical dimension of the space, the dimension of the order
parameter and its symmetries, the (short/long) range of interaction of the coupling.
A universality class is uniquely defined by its critical exponents,
which describe the behaviour of the order parameter in proximity of
the critical point as a function either of the control parameter
(e.g. the temperature) or of the conjugate field of the
order parameter (e.g. the magnetic field).

When passing to nonequilibrium processes, 
the phenomenology is much wider and a classification in universality classes is not as firm.
A much studied case is the so-called ``phase separation". Let us consider
a system undergoing a continuous phase transition (at $T=T_c$) when passing from
a disordered high temperature phase to an ordered low temperature phase.
If the temperature $T$ is suddenly decreased (quenching) from $T_i>T_c$ to $T_f<T_c$,
the system undergoes an ordering process where the typical size $L$ of
ordered regions increases in time, $L(t)$.
This process, called coarsening, lasts forever (for infinite systems) if the system
is globally at thermodynamic equilibrium.
In most cases $L(t)$ increases as a power law, $L(t)\sim t^n$, which defines
the coarsening exponent $n$.
Generally speaking, $T_i$ and $T_f$ are irrelevant parameters and it
appears that the physical space dimension is also irrelevant (as long as
$T_c$ is finite, $T_c > 0$). It appears instead that conservation laws are relevant for the
dynamics and it is reasonable to expect that a conservation law slows down the dynamics
and reduces the coarsening exponent, a known fact at present \cite{Bray,*Bray_Rutenberg}.

The spirit of universality also means that the same model is important
for different physical problems: for example, phase separation and pattern
formation have several similarities.
Therefore our study, which focuses on a certain
class of growth equations for crystals exhibiting pattern formation, is expected to be relevant for both fields.
This class of equations, see Eq.~(\ref{eq.crescita}), has emerged in the last twenty years as a prototypical
description of crystal growth by deposition processes.
It has some similarities with well known models
as the Cahn-Hilliard equation and the clock models~\footnote{The clock models,
or vector Potts models, are 2D planar spin models where the spins are restricted to
$q$ evenly-spaced orientations}, but its
general properties have been now fully established, as discussed in the next section.

This equation leads to morphological instability of a planar surface, with formation of mounds/pyramids out
of the flat front. In general, as previously sketched for domains in phase separation processes, mounds coarsen, but under some conditions we show that other scenarios take place.
In apparent contrast to some existing literature (see Section~\ref{Sec_disc}),
we are able to state that
pattern symmetry is irrelevant and only two universality classes result,
depending on whether  mound's slope is constant (faceting) or it
 is an increasing function of time. This last feature is known a priori, from
visual inspection of the surface current $\bj(\bm)$ (see below) and allows the definition of a second exponent $\beta\ge 0$, describing the behavior of the typical mound's slope $M$ in time, $M(t)\sim t^\beta$.
The two universality classes we have found are therefore given, in the case of constant slope $\bm^*$ ($\bj(\bm^*)=0$), by $\beta=0$ and $n=1/3$, and, in the case of increasing slope, by $\beta > 0$ and $n=1/4$.

The idea used to establish these results is based on the statement that coarsening takes place
if the steady-state periodic solutions are unstable against perturbations of the phase of the pattern \cite{1d}. More precisely, a periodic pattern has a constant wavenumber $\mathbf{q}$
which acquires a space-time dependence when the pattern is perturbed
(we can also define the phase of the pattern):
if the periodic pattern is perturbed, the wavenumber (as well as the phase) will vary from one point to another.
If the perturbation grows with time, we say that the pattern is unstable with respect to wavenumber (or phase) fluctuations. 
If the periodic pattern is unstable with respect
to phase fluctuations then we expect coarsening to take place.
It will be shown that the phase of the pattern obeys a diffusion equation and instability is signaled by a negative diffusion coefficient $D$.
This diffusion coefficient (actually in two dimensions there are several diffusion coefficients, as we shall see) 
depends on the steady-state pattern properties,
and more particularly on the modulus of the wavenumber $q$. By using a dimensional relation,
$|D(q)|\approx {L^2}/{t}$, where $q=2\pi/L$, we shall extract the coarsening exponent.

Here we are able to make stronger and more general
statements with respect to \cite{prl_Sofia}, facing a wider range of two-dimensional patterns and stressing on universal features of unstable crystal growth.
This is the focus of the present paper: going beyond the details of
the equation and of the physical process and pointing out what is relevant, slope selection or not,
and what is not, the symmetry of the pattern and that of the mass current. Although no complete proof about physical space dimensionality is accomplished, our study reveals strong support regarding its irrelevance.
\newline

\section{Crystal growth equation}

In this Section we shall give a brief introduction to the class of equations 
we are interested in, mainly addressing
the qualitative aspects of the dynamics rather than their physical derivation (for a thorough discussion on the physical background, the reader is referred to \cite{Politi_review}).

A growing planar crystal  surface (growing by molecular beam epitaxy, for example) can undergo a morphological instability resulting into 
the formation of three-dimensional mounds or pyramids of linear
size $L$ and height $H$.
The subsequent morphological evolution may range from a 
pattern of constant 
$L$ and an increasing $H$ up to a
perpetual increase  of $L$ in the course of time (coarsening), 
with $H$ increasing in concert.
An intermediate scenario may also take place in some cases, 
where $L(t)$ increases up to a length
$L_{max}$ reached at a given time, beyond which the mound size is frozen, while 
mound height keeps growing. 
This scenario corresponds to interrupted coarsening \cite{Danker}. 
We are not aware of a scenario where both $L$ and $H$ keep constant in time.

In the case  of a perpetual coarsening the generic evolution law of 
$L(t)$ is algebraic with
coarsening exponent, $n$, defined as $L(t) \sim t^n$.
During the coarsening process, the typical slope $M\approx H/L$ 
may either keep constant
or increase in time, $M(t)\sim t^\beta$, therefore defining a second exponent $\beta\ge 0$.

From a mesoscopic point of view, the local velocity $\partial_t z$
of a surface $z(\mathbf{x},t)$ growing under a deposition flux 
of intensity $F_0$ must have the form
\be
\partial_t z(\mathbf{x},t) = F_0 - \nabla \cdot \mathbf{J}_{tot} ,
\ee
provided that  the deposited mass on the surface does not evaporate and that no holes occur in the growing solids~\footnote{More precisely, we should require that volume is conserved,
which forbids voids and overhangs.}.
The total current $\mathbf{J}_{tot}$ is a function of the slope $\mathbf{m}=\nabla z$  and higher
order spatial derivatives and it accounts for all surface rearrangement processes.
Its simplest form is
\be
\label{j2D}
\mathbf{J}_{tot} = \mathbf{j}(\mathbf{m}) + \Gamma \nabla^2 \mathbf{m} ,
\ee
where $\mathbf{j}(\mathbf{m})$ is a function of  the slope only and it accounts for the existence
of a mass current on a terrace. At small slopes
$\mathbf{j}\approx \nu \nabla z$: if the current is uphill ($\nu > 0$),
the flat surface is destabilized at sufficiently large scales.
The second term, $\nabla^2 \mathbf {m}$, regularizes the dynamics at short length scales and
it may have different physical origins~\footnote{Thermal detachment from steps
and fluctuations in the diffusing current are two of them.}.

By performing the substitution $z \rightarrow h=z-F_0 t$ and after appropriate  rescaling of $\mathbf{x}$ and $t$, it is possible
to absorb $\Gamma$ and $\nu$ into the new variables so that the equation can be written in the form
\begin{equation}
\label{eq.crescita}
\frac{\partial h(\mathbf{x},t)}{\partial t} = -\mathbf{\nabla}\cdot {[\mathbf{j}(\nabla h) +
\mathbf{\nabla}(\nabla^2 h)]} \equiv -\mathbf{\nabla}\cdot {\mathbf{J}_{tot}},
\end{equation}
where $\bj(\nabla h)=\nabla h +$ {\tt higher order terms}.

Some important features of the nonlinear dynamics can be discussed by referring to the
one dimensional version of Eq.~(\ref{eq.crescita}), which has been discussed at length
in Ref.~\cite{1d}:
\be
\label{eq_1d}
\partial_t h = -\partial_x [ j(h_x) + h_{xxx} ] .
\ee
In fact, by taking the spatial derivative of both sides, we get the generalized Cahn-Hilliard equation,
\be
\partial_t m = -\partial_{xx} [ j(m) + m_{xx} ]
\ee
where the shape of the potential $U(m)=\int dm j(m)$ determines the type of
dynamics~\footnote{$U(m)$ is an even potential with a minimum in $m=0$, since $j(m)= m$ at small
$m$.}:
(i)~stationary solutions, satisfying $j(m) + m_{xx}=0$, correspond to periodic ``oscillations'' within the potential
well $U(m)$;
(ii)~there is coarsening if and only if the wavelength of such stationary solutions is an increasing function
of their amplitude; in general three scenarios, depicted here above, are possible:
perpetual coarsening, interrupted coarsening, no coarsening;
(iii)~the slope of emerging mounds is constant if $U(m)$ has maxima at finite $m=\pm m^*$,
otherwise slope increases forever.

When passing from one to two dimensions, i.e. passing from Eq.~(\ref{eq_1d}) to
Eq.~(\ref{eq.crescita}), the equivalence between the growth equation and the
Cahn-Hilliard equation ceases to be valid~\cite{Politi_review} (see also Section~\ref{sec_discussion}).
Furthermore, the surface current $\mathbf{j}$ requires specification of its in-plane symmetry,
which adds a new  degree of freedom to the problem.
The following Sections present the various dynamical scenarios where the values of the exponents $n$ and
$\beta$ (universality classes) are extracted
for the family of models defined by Eq.~(\ref{eq.crescita}).
We shall follow  a multiscale perturbative approach, discussed in the next Section,
which allows us to write down the phase diffusion equation that describes the evolution of the
typical mound size in the course of time. The various dynamical scenarios will constitute the subject  of
Sec.~\ref{sec_scenarios} while distinct universality classes will presented in Sec.~\ref{sec_classes}.
A thorough discussion of our results will follow in Sec.~\ref{sec_discussion}.

\section{The phase diffusion equation}
\label{sec_pde}

As already anticipated in the previous Section, the flat profile, namely the solution $h \equiv 0$ of Eq.~(\ref{eq.crescita}), is unstable. This is easily shown
from a linear stability analysis: setting $h=\delta~ \hbox{exp}(\omega t + i \mathbf{k} \cdot \mathbf{x})$ in Eq.~(\ref{eq.crescita})
and assuming $\delta \ll 1$, we obtain the linear spectrum:
\be \label{spectrum}
\omega(k)=k^2-k^4 ,
\ee
where $k=|\mathbf{k}|$. This result  shows that there is a band of  wave-vectors ($0<k<1$) with  positive $\omega$, so the corresponding harmonic amplitude
increases exponentially with time until nonlinearities can no longer be disregarded. This instability will result first
in a deformed (more or less regular) surface and during the initial stages the amplitude
grows quite rapidly. 

Interesting nonlinear dynamics appears later and periodic steady-state solutions
play the major role, because relevant informations can be drawn from their stability%

The general idea used here is that if coarsening takes place, this means that every steady-state solution is unstable with respect to wavelength fluctuations and therefore the relevant variable to describe
this phenomenon is the wavelength, or, more precisely, the phase of the pattern, since in nonlinear systems it is known that the phase is a more appropriate variable to deal with rather
than the wavelength itself \cite{bender_orszag_book}. 
This idea was applied with success to study one dimensional fronts in Ref.~\cite{1d},
where the ability of the system to develop coarsening was directly related to steady-state
properties, with
no need to perform a forward time-dependent calculation. It will even be shown for several examples below, that the stability or instability of the pattern against phase fluctuations can be concluded analytically.
Even more importantly, our approach provides the values of exponents $n$ and $\beta$.

In order to study stability of the periodic steady-state $h_0$, we seek for
solutions of the nonlinear equation in the form (with $\varepsilon$ small parameter)
\be \label{exph} h=\th_0+\varepsilon \th_1+\dots ,
\ee
and linearize the equation. However,
in addition to this quite standard study of linear stability, the crux of our method is to introduce a multiscale analysis that will allow us to extract the phase evolution equation,
the analysis of which will inform us on the presence of coarsening or the lack thereof.
Therefore, besides the fast variables $\mathbf{x}$ and $t$, we introduce slow variables defined as
\be \label{espansioni}
\mathbf{X} = \varepsilon \mathbf{x}  \, , ~~ T= \varepsilon^2 t .
\ee
The perturbation parameter
$\varepsilon$ is a small quantity that defines the fact that we are looking for long wavelength modulation of the pattern, which are the most \textquotedblleft dangerous\textquotedblright modes (see \cite{1d} for more details).
In a multiscale
spirit fast and slow variables are treated as if they were  independent~\cite{Hoyle}.
As already said, it is convenient to work with the phase variables rather than with the
spatial variables. For that purpose, we introduce (in two dimensions) two scalar  phase variables $\varphi_1$ and $\varphi_2$. If the pattern is perfectly periodic then these variables are simply given by
\begin{equation}\label{phi}
\varphi_1 := \mathbf{q}_1\cdot\mathbf{x} ,\quad \varphi_2 :=\mathbf{q}_2\cdot\mathbf{x} ,
\end{equation}
where $\mathbf{q}_i=\nabla \varphi_i$ are the basis wave vectors defining the symmetry of the stationary periodic pattern.
To account for perturbations of the periodic lattice, $\mathbf{q}$-vectors are not just constants but have a dependance on slow scales: $\mathbf{q}=\mathbf{q}(T,\mathbf{X})$;
therefore we introduce for convenience the slow phase scales: $\psi_{i}=\varepsilon \varphi_{i}$, so that $\mathbf{q}_{i}=\nabla_{\mathbf{X}} \psi_{i}$ can be expressed as function of
slow variables only.

According to this approach, various differential operators in the model equation have to be substituted as follows:
\bea
\label{time}
\partial_t \rightarrow \varepsilon^2 \partial_T &=&
\varepsilon^2 [(\partial_T \varphi_1) \partial_{\varphi_1} + (\partial_T \varphi_2) \partial_{\varphi_2} ]\nonumber \\
&=&
\varepsilon [(\partial_T \psi_1) \partial_{\varphi_1} + (\partial_T \psi_2) \partial_{\varphi_2}] , \\
\label{space}
\nabla \rightarrow \nabla_0 + &\varepsilon \nabla_\mathbf{X}& ,
\eea
with $\nabla_0 = \mathbf{q}_1 \partial_{\varphi_1} + \mathbf{q}_2 \partial_{\varphi_2}$ and $\nabla_{\mathbf{X}} = (\partial_X, \partial_Y)$.
Then expansions (\ref{exph}), (\ref{time}) and (\ref{space}) are reported into the model equation (\ref{eq.crescita}) which yields (by keeping only terms up to order $\varepsilon$, see
Appendix~\ref{operators})
\be
\begin{split}
\varepsilon & [(\partial_T \psi_1) \partial_{\varphi_1} \th_0 + (\partial_T \psi_2) \partial_{\varphi_2} \th_0]
= \\
= & -(\nabla_0+\varepsilon \nabla_{\mathbf{X}}) \cdot \{  \mathbf{j}(\nabla_0 \th_0) +\varepsilon \mathcal{J} (\nabla_0 \th_1 + \nabla_\mathbf{X} \th_0) \\
+ & \nabla_0 (\nabla_0^2 \th_0) + \varepsilon [\nabla_\mathbf{X} (\nabla_0^2 \th_0) + \nabla_0 (\nabla_1^2 \th_0 + \nabla_0^2 \th_1)] \} ,
\end{split}
\ee
to be studied order by order.

{\it Zeroth order} --
The zeroth-order defines stationary solutions $\th_0$ as the unperturbed ones:
\begin{equation} \label{ord.zero}
0 = \nabla_0 \cdot {[\mathbf{j}(\nabla_0 \th_0) + \nabla_0(\nabla_0^2 \th_0)]}
= \nabla_0 \cdot {(\mathbf{J}_0)_{tot}}
\equiv \mathcal{N}[\th_0] ,
\end{equation}
where $\mathcal{N}$ is a nonlinear operator acting on $\th_0$. Explicit solutions $\th_0$ are in general not 
available, the only basic information being that  $\th_0$
 enjoys  periodicity properties in $\varphi_1$, $\varphi_2$.
Focusing our analysis on high symmetry substrates, for which $\langle h\rangle =0$, a stronger condition can be imposed:
\be \label{corr_tot}
(\mathbf{J}_0)_{tot}=0 .
\ee

{\it First order} --
At first order we obtain a linear and inhomogeneous equation for $\th_1$:
\be \label{ord.uno}
\mathcal{L}[\th_1] = g(\th_0,\psi_1,\psi_2),
\ee
where
\bea
\label{expg}
g &\equiv & (\partial_T \psi_1) \partial_{\varphi_1} \th_0 + (\partial_T \psi_2) \partial_{\varphi_2} \th_0
+ \nabla_0 \cdot [\mathcal{J}(\nabla_{\mathbf{X}} \th_0) \nonumber \\
&+&\nabla_{\mathbf{X}}(\nabla_0^2 \th_0) + \nabla_0(\nabla_1^2 \th_0)]
\eea
is a function of stationary solutions $\th_0$, while
\be
\label{linear}
\mathcal{L}[\th_1] \equiv -\nabla_0 \cdot {[\mathcal{J}(\nabla_0 \th_1) + \nabla_0 (\nabla_0^2 \th_1)]}
\ee
is the Fr\'{e}chet derivative of $\mathcal{N}$, defined as
\be
\mathcal{N}[\tilde h_0+\varepsilon \tilde h_1]=\mathcal{N}_0[\tilde h_0]+ \varepsilon \mathcal{L}[\tilde h_1].
\ee
By virtue of translational invariance of $\mathcal{N}$ with respect to space variables, it follows that
$\mathcal{N}[\th_0(\varphi_i+\Delta_i)]$ must vanish as well.
In the limit $\Delta_i\to 0$ we get
\be
\mathcal{N}[\th_0(\varphi_i+\Delta_i)] =
 \mathcal{N}[\th_0(\varphi_i)] + \Delta_i \mathcal{L}[\partial_{\varphi_i} \th_0(\varphi_i)]=0 ,
\ee
which also implies that $\mathcal{L}[\partial_{\varphi} \th_0(\varphi)]=0$. 
Therefore, since $\mathcal{L}[\th_1] =0$ has nontrivial solutions ($\partial_{\varphi} \th_0$), the Fredholm alternative theorem~\cite{zwillinger} can be used for Eq.(\ref{ord.uno}).
Such theorem guarantees solutions for Eq.~(\ref{ord.uno}) if and only if the so called {\it solvability conditions} (expressing the fact that the right hand side of Eq.~(\ref{ord.uno}) is orthogonal to the kernel of the adjoint operator of $\mathcal{L}$) are verified.
These conditions have the following form
\footnote{The scalar product is defined as $
\protect\langle f,g \protect\rangle
:= \frac{1}{(2\pi)^2} \int_0^{2\pi} \int_0^{2\pi} d\varphi_1d\varphi_2 f^* g$.}:
\begin{equation} \label{cond.risolv}
\langle v_i,g\rangle =0 ,
\end{equation}
where functions $v_1$, $v_2$ verify ${\cal L}^\dagger [v]=0$.
We therefore calculate the adjoint ${\cal L}^\dagger$ of our linear operator from the definition $\langle \mathcal{L^{\dag}}v,u \rangle = \langle v,\mathcal{L} u \rangle$.
Given that
\begin{equation}
\mathcal{L}[u]=-\nabla_0 \cdot \mathcal{J}\nabla_0 u - \nabla_0^2 (\nabla_0^2 u),
\end{equation}
$\mathcal{L}$ is self-adjoint if and only if the Jacobian matrix ${\cal J}$ is symmetric (see Appendix~\ref{adjoint}).
This latter case is definitely the most common one, since we find that it is assured by all the explicit forms of $\mathbf{j}$ used in the literature.
We also stress that a symmetric ${\cal J}$ means that the current derives from a potential, $\bJ=-\delta {\cal F}/\delta\bm$
(see Section\ref{sec_discussion} for further details).

For the sake of completeness, we must keep in mind that the symmetry
property for $\mathcal{J}$ is not a limit in applicability of the current method: a phase diffusion equation could be derived formally without having a linear self-adjoint operator.
However, for non adjoint operators, the solutions of 
$\mathcal{L}^\dag [v]=0$ can be obtained, in general, only numerically \cite{matteo},
even if examples to get them analytically in $1d$ do exist~\cite{1d}. 

According to the above discussion, 
if ${\cal J}$ is symmetric than ${\cal L}={\cal L}^\dag$ and $v_i=\partial_{\varphi_i} \th_0(\varphi_i)$.
It is now possible to rewrite $g$, see Eq.~(\ref{expg}),
as follows (see Appendix~\ref{c_abg} for more details):
\begin{equation}
g \equiv (\partial_T \psi_1) \partial_1 h_0 + (\partial_T \psi_2) \partial_2 h_0 - (\psi_{\alpha})_{\beta\gamma} c^{\alpha}_{\beta\gamma},
\end{equation}
where 
$\partial_{\alpha}\equiv \partial_{\varphi_\alpha}$ and $h_0\equiv \th_0$ 
for ease of notation, and
\bea
-c^{\alpha}_{\beta\gamma} &=& q_{\delta\nu}\partial_{\delta}
\left[\mathcal{J}_{\nu\gamma}\frac{\partial \th_0}{\partial q_{\alpha\beta}}\right]
+ 2 q_{j\gamma}q_{l\beta}\partial_{\alpha} \partial_l \partial_j \th_0
\nonumber \\
&+& 3 \nabla_0^2 q_{\nu\beta} \partial_{\nu} \frac{\partial \th_0}{\partial q_{\alpha\gamma}}
+ \delta_{\beta\gamma} \nabla_0^2 \partial_{\alpha} \th_0 ,
\label{c_generali}
\eea
with $q_{ij}=(\mathbf{q}_i)_j$ as the $j$-th component of the $i$-th wave-vector.
Moreover, the compact notation $h_j \equiv \partial_{\varphi_j}\th_0$ will be adopted from now on.

By using the above expression for $g$ in the two solvability conditions
(\ref{cond.risolv}) we obtain
 the phase diffusion equations ($i=1,2$):
\be \label{eq.diff.gen}
\partial_T \psi_i = \frac{\partial \psi_{\alpha}}{\partial X_{\beta} \partial X_{\gamma}}
\tilde D_{\beta\gamma}^{i\alpha} ,~~~~~\alpha,\beta,\gamma=1,2
\ee
where repeated indices are to be summed over according to Einstein's convention.
The diffusion coefficients have the following expressions:
\be
\label{diffusion_coeff}
\tilde D_{\beta\gamma}^{1\alpha}=\left[\frac{\langle h_1,c^{\alpha}_{\beta\gamma}\rangle \langle h_2,h_2\rangle  -
\langle h_2,c^{\alpha}_{\beta\gamma}\rangle \langle h_1,h_2\rangle }{\langle h_1,h_1\rangle \langle h_2,h_2\rangle -{\langle h_1,h_2\rangle }^2}\right]
\ee
and $\tilde D_{\beta\gamma}^{2\alpha}~ \substack{{1 \leftrightarrow 2} \\ \displaystyle =}~
\tilde D_{\beta\gamma}^{1\alpha}$.
It is convenient to define new diffusion coefficients $D^{i\alpha}_{\beta\gamma}$, by regrouping similar derivatives:
\be \label{convenz.coeff}
D^{i\alpha}_{\beta\gamma} = \left\{
\begin{array}{cc}
\tilde D^{i\alpha}_{\beta\gamma} & \beta = \gamma \\
\tilde D^{i\alpha}_{\beta\gamma} + \tilde D^{i\alpha}_{\gamma\beta} & \beta \neq \gamma.
\end{array}
\right.
\ee
Therefore, in the most general case, Eqs.~(\ref{eq.diff.gen}) have twelve independent
diffusion coefficients.
Their expressions are in general quite involved except if some symmetry properties of the steady-state solutions $h_0$ are evoked. Symmetry properties will lower the number  of independent diffusion coefficients.
It should be remembered  that
$h_0$ is a perfectly periodic in-plane pattern, defined by one of the five known two-dimensional Bravais lattices.
Selecting one of these patterns for the stationary solution $h_0$ means fixing the two $\mathbf q$-vectors and the space group symmetry
that leave $h_0$ unchanged.
It is convenient to list the Bravais lattices in a sort of hierarchy to face at once how the demand of symmetry simplifies the expression of diffusion equations.
Let us define $\Theta$ as
the angle between the two $\mathbf q$-vectors and $p$, the proportionality between their moduli:
\be
\cos\Theta := \frac{\mathbf{q}_1 \cdot \mathbf{q}_2}{|\mathbf{q}_1||\mathbf{q}_2|} \, , \quad p :=\frac{|\mathbf{q}_1|}{|\mathbf{q}_2|} \, .
\label{Theta_p}
\ee
The proposed order for the five Bravais lattices is shown in Table I.
\begin{table}[t]
\begin{tabular}{c | c | c | c | c}
symmetry of $h_0(\mathbf{x})$ & $\Theta$ & $p$ & invariances & $D^{i\alpha}_{\beta\gamma}$\\
\hline
oblique & no specific & no specific & 2-fold & 12\\
rhombic & no specific & 1 & 2-fold, $\Pi_2$ & 6\\
rectangular & $\pi/2$ & no specific & 2-fold, $\Pi_2$ & 6\\
square & $\pi/2$ & 1 & 4-fold, $\Pi_2$ & 3\\
hexagonal & $\pi/3$ & 1 & 6-fold, $\Pi_2$ & 2\\
\hline
triangular & $2\pi/3$ & 1 & 3-fold, $\Pi_1$ & 2\\
\end{tabular}
\caption{Presentation of the 5 two dimensional Bravais lattices, classified according to relative orientations between the two $\mathbf{q}$-vectors ($\Theta$) and their relative amplitude ($p$);
moreover, parity symmetry ($\Pi_2$ with respect to both space variables, $\Pi_1$ with respect to a 
single space variable) and rotational invariances are specified. The last, extra, row refers to the 3-fold case.
An increasing symmetry corresponds to a decreasing number of independent diffusion coefficients $D^{i\alpha}_{\beta\gamma}$.
}
\end{table}
In addition, given its considerable relevance to experiments \cite{Pt111} we also studied the 3-fold case, that is not included among the Bravais lattices but, nevertheless,
can be dealt with using the same method as for the other symmetries. The 3-fold case is characterized by
$\Theta= 2\pi/3$ and a $p=1$, while the parity symmetry holds for a single space variable only.

In Appendix~\ref{hex.diff} we provide an  explicit treatment of the phase diffusion equation for the hexagonal symmetry and determine the
number of independent coefficients. This serves as a guide for the other symmetries for which we do not report 
the details. Our results are summarized in Table I:
last column reports the number of independent coefficients $D^{i\alpha}_{\beta\gamma}$ corresponding to each pattern symmetry.
For oblique, that is the most general one, the number of independent $D^{i\alpha}_{\beta\gamma}$s is in fact twelve; for rhombic and rectangular ones parity allows to reduce this number to six;
then, the increased degree in the rotational invariance for the square and hexagonal cases implies further reduction to, respectively, three and two independent coefficients.
The 3-fold symmetry shares similarities with  the hexagonal pattern (albeit the two symmetries are distinct). It turns out that these two symmetries obey formally the same diffusion equation, with the same number of independent coefficients.

In the next Section we are going to exploit the phase diffusion equations for some symmetries and we will report on some far-reaching consequences. In particular, we will
examine stability of Eq.(\ref{eq.diff.gen}) with respect to phase perturbations, a relevant information regarding the coarsening problem.

\section{The coarsening conditions}
\label{sec_scenarios}

A coarsening dynamics is signaled by phase instability, i.e. by
a phase which increases exponentially with time~\footnote{This is strictly true for short times, 
when Eq.~(\protect\ref{eq.diff.gen}) is exact.}.
Phase diffusion equations, Eqs.~(\ref{eq.diff.gen}), are linear and can  be solved assuming
\be
\label{phase}
\psi_{1,2}(\mathbf{X},T)=\psi_{1,2}^{(0)} \exp{(\Omega T)}\exp{(i\mathbf{K}\cdot\mathbf{X})}
\ee
and imposing a null determinant
for the linear system with unknowns $\psi_{1,2}^{(0)}$. This way,
we can write down a quadratic equation for $\Omega$
\be
\Omega^2+f(D_{\beta\gamma}^{i\alpha},\mathbf{K})\Omega + g(D_{\beta\gamma}^{i\alpha},\mathbf{K})=0
\ee
and obtain two entire spectra,
$\Omega_{1,2}=\Omega_{1,2}(\mathbf{K})$, whose properties depend on the symmetry of $h_0$.
We present here below detailed results regarding  rectangular, square, hexagonal and triangular symmetries. 
Appendix~\ref{q_vectors} lists the $\mathbf{q}$-vectors used in these specific cases.
The oblique and rhombic symmetries will not be treated here since they involve quite lengthy expressions. 
Since we do not expect any new specificity associated with them (see later
discussion), we did not feel it worthwhile to dwell on this issue.

\subsection{The hexagonal and triangular symmetries}
In the 6-fold and in the 3-fold cases the spectrum turns out to be isotropic in $\mathbf{K}$ and the two eigenvalues are found to be:
\begin{equation}
\Omega_1(K)=-D_{22}K^2 , ~~ \Omega_2(K)=-D_{11}K^2 ,
\end{equation}
where $D_{11}^{11} \equiv D_{22}^{22} \equiv D_{11}$ and $D_{22}^{11} \equiv D_{11}^{22} \equiv D_{22}$ (see Appendix~\ref{hex.diff}).
Since $D_{22}$ is positive:
\be
D_{22}= \frac{9q^2}{\langle h_{\varphi}^2 \rangle } \langle h_{12}^2 \rangle >0 ,
\ee
the eigenvalue $\Omega_1$ is negative, signaling stability of the pattern. The other eigenvalue, instead, has
no a priori fixed sign:
\begin{equation} \label{D.esag}
D_{11}=\frac{4q^{7/4}}{\langle h_1^2\rangle } \partial_q (q^{5/4}\langle h_{12}^2\rangle ).
\end{equation}
A negative $D_{11}$ would signal instability. We will see later
how to determine this sign analytically and how to discriminate among different dynamical scenarios.

\subsection{Square and rectangular symmetries}

For these symmetries, the spectrum of eigenvalues is anisotropic and its analysis is, in principle, more complicated. 
Let us first consider the square case spectrum:
\begin{widetext}
\be
\label{Omega_sq}
\Omega_{1,2}(\!K,\!\theta)\!=\! -\frac{K^2}{2}\Bigg[\! (D_{11}\!+\!D_{22}\!)\! \pm \!
\sqrt{\!(D_{11}\!-\!D_{22})^2 \!+\! 4[D_{12}^2 \!-\! (D_{11}\!-\!D_{22})^2]\sin^2(\theta)\! +\! 4[(D_{11}\!-\!D_{22})^2\!-\!D_{12}^2]\sin^4(\theta)} \Bigg] ,
\ee
where $K_1=K\cos\theta$, $K_2=K\sin\theta$
and where we have used the compact notations:
$D_{11}^{11} \equiv D_{22}^{22} \equiv D_{11}$, $D_{22}^{11} \equiv D_{11}^{22} \equiv D_{22}$,
and $D_{12}^{12} \equiv D_{12}^{21} \equiv D_{12}$.
Expression (\ref{Omega_sq})
shows that extremal values  for $\Omega_{1,2}(K,\theta)$ in the $(K_1,K_2)$ plane
are along the directions
$\displaystyle \theta=n \frac{\pi}{2}$ and $\displaystyle \theta=\frac{\pi}{4}+n\frac{\pi}{2}$.
Since we are dealing with a 4-fold symmetry, we consider just two cases, for each of which we distinguish two different eigenvalues:
\begin{itemize}
  \item $\theta=0$, $\Omega_1^0 (K) = -D_{22} K^2$ , $\Omega_2^0 (K) = -D_{11} K^2$;
  \item $\theta=\pi/4$, $\Omega_1^{\pi/4} (K) = -(D_{11}+D_{22}-D_{12}) K^2/2$ , 
$\Omega_2^{\pi/4} (K) = -(D_{11}+D_{22}+D_{12}) K^2/2$.
\end{itemize}

As already seen for hexagonal and triangular symmetries, also in this case 
one eigenvalue for each couple is always negative, since:
\begin{gather}
D_{22}= \frac{q^2}{\langle h_{\varphi}^2\rangle } [\langle h_{11}^2\rangle  + 3\langle h_{12}^2\rangle ]>0, \\
D_{11}+D_{22}-D_{12}= \frac{4q^2}{\langle h_{\varphi}^2\rangle } \langle h_{11}^2\rangle >0,
\end{gather}
while the sign of the other eigenvalues is not obvious, being determined by that of the following expressions:
\begin{gather} \label{autov.quad1}
D_{11} = \frac{1}{\langle h_1^2\rangle } [\partial_q (q^3 \langle h_{11}^2\rangle ) + q^3 \partial_q \langle h_{12}^2\rangle + q^2 \langle h_{12}^2\rangle  ] , \\
\label{autov.quad2}
D_{11} + D_{22}+D_{12} = \frac{4}{\langle h_1^2\rangle } \left[ \frac{1}{2} q^3 \partial_q \langle h_{11}^2\rangle  + q^2 \langle h_{11}^2\rangle 
+ \frac{1}{2} q^3 \partial_q \langle h_{12}^2\rangle + 2q^2 \langle h_{12}^2\rangle \right].
\end{gather}
In Section V we propose
calculations of the diffusion coefficients valid in the weakly nonlinear regime, in order to treat those eigenvalues
whose sign has not been easily recognizable. 

Analogously, for the rectangular case the spectrum takes the following form:
\be
\label{Omega_rec}
\begin{split}
&\Omega_{1,2}(K,s) = -\frac{K^2}{2} \Bigg\{ \Bigg[(D^{11}_{11}+D^{22}_{11})+(D^{11}_{22}+D^{22}_{22}-D^{11}_{11}-D^{22}_{11}) s \Bigg] \\
& \pm \! \sqrt{\! (D^{11}_{11}\!-\!D^{22}_{11})^2 \!+\! \Bigg[2 (D^{11}_{11}\!-\!D^{22}_{11})(D^{11}_{22}\!-\!D^{22}_{22}\!-\!D^{11}_{11}\!+\!D^{22}_{11}) \!+\!
4 D^{12}_{12}D^{21}_{12} \Bigg]\! s \!+\! \Bigg[\!(D^{11}_{22}\!-\!D^{22}_{22}\!-\!D^{11}_{11}\!+\!D^{22}_{11})^2
-4 D^{12}_{12} D^{21}_{12} \Bigg] s^2} \Bigg\}
\end{split}
\ee
\end{widetext}
where $s \equiv \sin^2(\theta)$.
The extremal values are now obtained not only along the maximal symmetry directions, 
namely along $\theta=n\pi$ and $\theta=n\pi \pm \pi/2$,
but also along two new other directions we are able to specify in the weakly nonlinear regime (see Appendix~\ref{rectangular}). 
According to such approximation, corresponding to steady states of small
amplitude $a$, these directions are
close to $\theta=\pi/4$:
\be
\sin^2(\theta)=\frac{1}{2} \pm 2 \sqrt{2} \frac{q^4}{m^2} \frac{p^2-1}{(p^2+1)^2} p \equiv \frac{1}{2} \pm O(a^2) ,
\label{otherext}
\ee
where $\displaystyle m= q^3 \partial_q (a^2)/a^2$ 
and $p$ is defined by Eq.~(\ref{Theta_p}).
The eigenvalues corresponding to the two first extremal directions are:
\begin{itemize}
  \item $\theta=0$ \\
$\Omega_1^0 (K) = -D_{11}^{22} K^2$ , \\
$\Omega_2^0 (K) = -D_{11}^{11} K^2$;
  \item $\theta=\pi/2$ \\
$\Omega_1^{\pi/2} (K) = -D_{22}^{11} K^2$ , \\
$\Omega_2^{\pi/2} (K) = -D_{22}^{22} K^2$.
\end{itemize}
Again, one eigenvalue for each couple is always negative, since:
\label{posit_rect}
\begin{gather}
D_{22}^{11}= \frac{q^2}{\langle h_1^2\rangle } [\langle h_{11}^2\rangle  + 3 p^2 \langle h_{12}^2\rangle ]>0, \\
D_{11}^{22}= \frac{q^2}{\langle h_2^2\rangle } [3 \langle h_{12}^2\rangle  + p^2 \langle h_{22}^2\rangle ]>0
\end{gather}
while the other has no obvious sign, as it is fixed by that of  the following expressions:
\be \label{autov.rect}
D_{11}^{11} = \frac{1}{\langle h_1^2\rangle } [ \partial_q (q^3 \langle h_{11}^2 \rangle)  + q^3 p^2 \partial_q \langle h_{12}^2\rangle + q^2 p^2 \langle h_{12}^2\rangle ] ,
\ee
\be
D_{22}^{22} = \frac{1}{\langle h_2^2\rangle } [p^2 \partial_q (q^3 \langle h_{22}^2\rangle) + q^3 \partial_q \langle h_{12}^2 \rangle  + q^2 \langle h_{12}^2\rangle ].
\ee

It is worth notice that $D_{22}^{11}=D_{11}^{22}$ and $D_{11}^{11}=D_{22}^{22}$, for $p=1$\footnote{Such a case, in fact, corresponds to congruent $q$-vectors, therefore (see Tab. I) to a geometry of the pattern that resembles the square one:
directions $\theta=0$ and $\theta=\pi/2$ must be equivalent high symmetry orientations, thus corresponding eigenvalues must be equal. Restoration of a square symmetry, however, is not complete because, for the present case, we have imposed a 2-fold invariance, not a 4-fold one.}.
The other two extremal directions, defined by Eq. (\ref{otherext}) have to be considered in the weakly nonlinear regime. The reader can find calculations in Appendix~\ref{rectangular}.
Here, it suffices to say that, also in the rectangular case, once the direction has been fixed, the sign of one eigenvalue is negative while the sign of the other is not evident.

As a summary of this section we can highlight two important conclusions. In the hexagonal and triangular symmetries, one eigenvalue is positive (phase instability) if the quantity
(see Eq.~(\ref{D.esag}))
\be
\label{a2d}
{\cal A} \equiv q^{5/4}\langle h_{12}^2\rangle
\ee
is a decreasing function of the wavenumber $q$. The quantity ${\cal A} $ depends only on the properties of the steady-state solutions.
Thus, determining whether coarsening occurs or not can be decided on the inspection of steady-state solutions only.
This result generalizes our previous one-dimensional study to two dimensions \cite{1d}, where we found that coarsening
occurs if $\langle h_{0}^2\rangle$ (which is nothing but the amplitude of the pattern) is a decreasing function of $q$.
In two dimensions~\cite{2d} we had previously found for the time-dependent Ginzburg-Landau
equation and for the Cahn-Hilliard equation that a certain quantity,
different both from that of the one-dimensional problem and from Eq.~(\ref{a2d}), must be a decreasing function of $q$.
Thus, we can state that the nature of the function whose decreasing character determines stability depends on the space dimension and on the class of the considered  equations
\footnote{For square and rectangular symmetries, we have not been able to find
explicitely this function. However, coarsening is still related to the negative sign of
a suitable diffusion coefficient.}.

The second important conclusion is that 
our results in this section do not depend on the nature of the current $\mathbf{j}$
entering in Eq. (\ref{eq.crescita}).
The reader can refer to Appendix~\ref{hex.diff} in order to check this statement in detail
for the 6-fold symmetry.

\section{The diffusion equation in the weakly nonlinear regime}

In this Section our aim is to analyze if 
coarsening occurs or not, while the determination of coarsening exponents will be presented in the next section.
In order to determine the dynamical scenarios for our growth equation, we need an evaluation of the signs of  appropriate diffusion coefficients, see
Eqs.(\ref{D11.esag},\ref{autov.quad1},\ref{autov.quad2},\ref{autov.rect}). This task can, in general, be performed only numerically by solving for the steady-state solutions. However, by restricting ourselves to a weakly nonlinear analysis, some analytical results can be obtained. To that end we assume that
 the amplitude of the stationary solution $h_0$ is small. 
We have already performed the general linear analysis of our equation,
which has resulted into the  spectrum (\ref{spectrum}).
In a weakly nonlinear approach we can push further this stability analysis extracting an approximated solution for $h_0$ in power series of the amplitudes of the Fourier modes.
Thanks to the periodic character of the stationary solution we can express $h(\mathbf{x},t)$ with a Fourier series that can
be truncated at some order.
The small amplitude limit is legitimate as long as    
$k \to  1$,
so that higher harmonics are stable, ensuring the consistency of the truncation of the series.

Since the symmetry of the growing pattern is identical or lower than substrate symmetry,
an isotropic current is the most general one, i.e. compatible with  any Bravais lattice.
We consider a generic class of isotropic currents $\mathbf{j}(\mathbf{m},c_2,c_4)=\mathbf{m}(1+c_2 \mathbf{m}^2+c_4 \mathbf{m}^4)$
so that Eq.~(\ref{eq.crescita}) becomes:
\be \label{wnsa}
\begin{split}
\partial_t h &= {\tilde L}[h]-c_2[3 (h_x^2 h_{xx} + h_y^2 h_{yy}) + h_y^2 h_{xx} + h_x^2 h_{yy} \\
&+\! 4 h_x h_y h_{xy}] \!-\! c_4 [5(h_x^4 h_{xx} \!+\! h_y^4 h_{yy}) \!+\! h_x^4 h_{yy} \!+\! h_y^4 h_{xx} \\ &+ 6h_x^2 h_y^2 (h_{xx}+h_{yy})+ 8 h_{xy}(h_x h_y^3 + h_x^3 h_y) ] ,
\end{split}
\ee
where ${\tilde L}[h]= -\nabla\cdot (\nabla h) - \nabla^4 h$ is the linear part.
We focus here on the  square and hexagonal symmetries for which the determination of the steady-state solution $h_0$ is relatively simple.

For a square symmetry, adopting the wave vector directions which  are specified in
  Appendix~\ref{q_vectors}, we can write
$h(\mathbf{x},t) = \sum_{n,m} a_{n,m}(t) e^{iq(nx+my)}$. The constant term ($n=0,m=0$) is zero because of the 
condition $\langle h \rangle =0$.
Given the real character of $h(\mathbf{x},t)$ and its parity-symmetry with respect to both space variables, it follows that
$a^*_{n,m}=a_{-n,-m}$ and $a_{-n,m}=a_{n,m}=a_{m,-m}$.
Invariance under $\pi/2$ rotations implies the following additional relations:
\begin{equation} \label{rotaz}
h(x,y)=h(-y,x)=h(-x,-y)=h(y,-x) ,
\end{equation}
which translate into the following conditions for the harmonic amplitudes:
\begin{equation} \label{coeff.Fourier.quad}
a_{n,m}=a_{m,-n}=a_{-n,-m}=a_{-m,n} .
\end{equation}
Therefore, at first order ($n,m=\pm 1$):
\begin{equation} \label{1armon}
h(\mathbf{x},t) = a_1(t) (e^{iq(x+y)}+ e^{iq(x-y)} + \hbox{c.c.}) ~,
\end{equation}
with $a_{1,1}=a_{-1,1}=a_{-1,-1}=a_{1,-1} \equiv a_1$, so the same dynamics equation holds for the four harmonics.
This series can be written in the more convenient form
$h_0=4 a_1 \cos \varphi_1 \cos \varphi_2$ and diffusion coefficients (\ref{autov.quad1}) and (\ref{autov.quad2}) can be explicitly calculated in the small amplitude limit:
\bea \label{wnsa_squared}
D_{11} &\simeq& \frac{4}{\langle h_{\varphi}^2 \rangle} \left[ \frac{d}{dq} (q^3 a_1^2) + q^3 \frac{d}{dq} a_1^2 + q^2 a_1^2 \right] \nonumber \\
&\simeq& \frac{8 q_{\hbox{\tiny MAX}}^3}{\langle h_{\varphi}^2 \rangle} \frac{da_1^2}{dq}  , \\
D_{11} &+& D_{22}+D_{12}\simeq \frac{16 q_{\hbox{\tiny MAX}}^3}{\langle h_{\varphi}^2 \rangle} \frac{da_1^2}{dq}  ,
\eea
with the substitution $q=q_{\hbox{\tiny MAX}}=1/\sqrt{2}$ in the last passages.
It is now evident how the sign of the eigenvalues is directly related to the increasing or decreasing character of the steady amplitude $a_1$ as function of the steady wave length $\displaystyle \lambda=2\pi/q$.
Reporting expansion (\ref{1armon}) into Eq.~(\ref{wnsa}) we find the equation obeyed by $a_1$:
\begin{equation} \label{a_1}
\dot{a}_1= a_1 (\omega_1 +c_2 20 q^4 a_1^2 + 224 c_4 q^6 a_1^4),
\end{equation}
where $\omega_1:=\omega(k=\sqrt{2}q)$, and then solve for stationary solutions:
\begin{equation} \label{altre}
a_1=0 , ~~ \omega_1 + 20 c_2 q^4 a_1^2 + 224 c_4 q^6 a_1^4 =0 .
\end{equation}
We conclude that the number of stationary solutions depends on values of $c_2$ and $c_4$: coarsening occurrence is directly associated with specific forms of $\mathbf{j}$ currents,
whose expression determines one of three possible scenarios.

Let's first consider the case in which $c_4 = 0$. We find:
\be \label{a}
a_1=\sqrt{\frac{\omega_1}{-20 c_2 q^4}} ~,
\ee
so that stationary solutions corresponding to the band of wavevectors such that $\omega_1>0$ exist only if $c_2<0$.
Using Eqs.~(\ref{wnsa_squared}), we now obtain:
\be
D_{11} = \frac{8}{5 \langle h_{\varphi}^2 \rangle} \frac{1}{c_2}  \; ,
\ee
Consequently, phase equation eigenvalues are positive, implying instability with respect to phase fluctuations, i.e. coarsening (this also holds for $c_2>0$ and $c_4<0$).
Instead, in the  case where $c_2>0$,  we can easily see that eigenvalues are negative, meaning no coarsening at all.
Finally, in the case $c_2<0$ and $c_4>0$, we find interrupted coarsening: the length scale $L$ of the pattern increases until
reaching  a certain maximum wavelength $\displaystyle \lambda=2\pi/q$. To fix idea, and without loss of generality, we set $c_2=-1$ and deduce the two stationary solutions from Eqs.~(\ref{altre}):
\begin{equation} \label{soluz.cr}
x_{\pm}(q,c_4)= \frac{10 q^4 \pm \sqrt{(10 q^4)^2-224 c_4 q^6 \omega_1}}{224 c_4 q^6} \equiv a_1^2 ,
\end{equation}
which both coincide at the maximum reachable length scale, where coarsening is interrupted.

For the hexagonal case, we proceed in the same way, setting a new truncated Fourier series:
\begin{equation} \label{1armon.hex}
h(\mathbf{x},t) = a_1(t) (e^{iq/2(x+\sqrt{3}y)}+ e^{iqx} + e^{iq/2(x-\sqrt{3}y)} +\hbox{c.c.}) ,
\end{equation}
with $a_{1,1}=a_{1,0}=a_{0,-1}=a_{-1,-1}=a_{-1,0}=a_{0,1} \equiv a_1$, so the same dynamics equation holds for any of the six harmonics
\footnote{For a general 6-fold Fourier series of kind $h(\mathbf{x},t) = \sum_{n,m} a_{n,m}(t) e^{iq/2[(n+m)x+\sqrt{3}(n-m)y]}$
we find the following relations:
\be \label{coeff.Fourier.hex}
\begin{split}
& a_{n,m}=a_{m,m-n}=a_{m-n,-n}= \\
& = a_{-n,-m}=a_{-m,-(m-n)}=a_{-(m-n),n}
\end{split}
\ee
to be valid among amplitudes.}.
Again, reporting Eq.(\ref{1armon.hex}) into Eq.(\ref{wnsa}), we obtain the amplitude equation:
\begin{equation} \label{a_1.hex}
\dot{a}_1= a_1 (\omega_1 +c_2 9 q^4 a_1^2 + 94 c_4 q^6 a_1^4),
\end{equation}
where $\omega_1:=\omega(k=q)$. Because this equation has the same structure and the same signs in front of each term as in Eq.  (\ref{a_1}), the same conclusions as above are reached, namely  we have coarsening,
no coarsening and interrupted coarsening scenarios depending on the signs of the coefficients $c_2$ and $c_4$.

The above results can be extended to other symmetries, including the 3-fold
symmetry, which does not correspond to a Bravais lattice. In that case, the starting
Fourier series corresponds 
to a linear combination of $h(\bx,t)$ and $h(\bx -\bx_0,t)$,
where $h(\bx,t)$ is given by Eq.~(\ref{1armon.hex}) 
and $\bx_0=(\frac{4}{3}\pi/q,0)$
\footnote{In order to get the 3-fold pattern, it is enough to suitably superpose
two 6-fold patterns. The value of $\bx_0$ can be easily understood as the
nearest neighbor distance in a honeycomb lattice, when this lattice is
considered as the superposition of two 6-fold lattices.}.

It is worthnoting that in the limit of
weak amplitude, the coarsening criterion always corresponds to the requirement that
the amplitude of the stationary solution is a decreasing function of the wavevector.
This is a trivial consequence of the single harmonic approximation, where
$h_0(\bx)=a_1(q)\times$ {\it exponential factors}.

\section{Universality classes}
\label{sec_classes}

In this section our aim is to extract analytically the coarsening exponents
$\beta$ and $n$, defined
as in Sec.~I.
In order to determine the exponent 
 we make use  of the temporal  behavior  
of the phase, see Eq.~(\ref{phase}), whose
amplitude increases as $\psi^{(0)}e^{\Omega T}$. The relevant time scale of the phase
instability is therefore set by $\Omega^{-1}$.
According to Sec.~\ref{sec_scenarios}, the unstable mode has an 
 eigenvalue of the form
$\Omega = -K^2{\cal D}$, where ${\cal D}$ is a suitable combination of diffusion
coefficients and whose negative sign indicates instability. Therefore, if $L$ is the
typical size of mounds after a time $t$, we have $T=\varepsilon^2 t$, $K\approx
\displaystyle \frac{1}{\varepsilon L}$ and
\be \label{dimensional}
|{\cal D}(q)| \approx \frac{L^2}{t},
\ee
with $q=2\pi/L$\footnote{It should be stressed that in the previous formula $L$ depends
on time because coarsening occurs, but the wavevector $q=2\pi/L$ appearing
in ${\cal D}(q)$ characterizes periodic stationary configurations}.

It turns out that the coarsening exponent $n$ only depends on 
one single property (see below) of the current $\bj$ entering the general equation (\ref{eq.crescita}),
while its symmetry, as well as the pattern symmetry, is definitely irrelevant. 
The only essential ingredient is whether the current leads or  not to a slope selection. We find 
 $n=1/3$ for $\mathbf{j}$ currents giving rise to mounds that grow with a certain constant slope and $n=1/4$ otherwise. Let us show more precisely these results.

Let us consider the square symmetry. We have seen that
coarsening occurs if at least one of the expressions given by (\ref{autov.quad1}) or (\ref{autov.quad2}) is negative.
Consider one scalar product entering expression (\ref{autov.quad1}):
\be \label{prod_scal}
\langle h_{11}^2 \rangle= \frac{1}{(2\pi)^2} \int_{0}^{\lambda}dx\int_{0}^{\lambda}dy \frac{1}{q^2}
\left( \frac{\partial^2 h_0}{\partial x^2}\right)^2 .
\ee
For systems exhibiting slope selection, the current $\mathbf{j}(\mathbf{m})$ has zeros for finite values $\mathbf{m}^*$ of the slope, therefore $\mathbf{m}^*=(\partial_x h_0,\partial_y h_0)$ is constant everywhere but along domain walls,
that have a finite but small thickness. Let's denote the thickness by $\delta$.  In the large 
wavelength limit $\delta \ll \lambda$  we can also assume that inside domain wall there is a  linear space dependence for the
slope $\mathbf{m}$: for example, in Eq.~(\ref{prod_scal}), $\partial h_0/\partial x = m_x \approx Ax +By$, with $A$ and $B$ real constants, whose exact values  are unimportant for our purposes.
Estimation of Eq.~(\ref{prod_scal}), therefore, yields
\be
\langle h_{11}^2 \rangle \simeq \frac{1}{(2\pi)^2}\frac{1}{q^2} \int_{0}^{\lambda} dy \int_{\lambda-\delta}^{\lambda}dx ~ A^2 \simeq c_{11} \lambda^3 + o(\lambda^3)
\ee
with $c_{11}$ a positive constant.
Similar considerations lead to $\langle h_{12}^2 \rangle \simeq c_{12} \lambda^3$ and
$\langle h_{1,2}^2 \rangle \simeq c_{\varphi} \lambda^2$, where constants, again, are positive.
We straightforwardly obtain $\displaystyle D_{11} = -2q c_{12}/c_{\varphi}$ from Eq.~(\ref{autov.quad1}) and $(D_{11}+D_{22}+D_{12}) =2q (c_{12}-c_{11})/c_{\varphi}$, with $c_{11} \geq c_{12}$\footnote{The sign of $D_{11}+D_{22}+D_{12}$
is determined by the inequality $c_{11}>c_{12}$, whose validation has to be brought back to the study of the two integrands $\displaystyle \left( \frac{\partial^2 h_0}{\partial x^2}\right)^2$ and $\displaystyle
\left( \frac{\partial^2 h_0}{\partial x\partial y}\right)^2$ appearing in the scalar products. For symmetry reasons, this is equivalent to
\be
\frac{1}{2}\left[ \left( \frac{\partial^2 h_0}{\partial x^2} \right)^2 + \left(\frac{\partial^2 h_0}{\partial y^2}\right)^2 \right] \ge
\left( \frac{\partial^2 h_0}{\partial x\partial y}\right)^2.
\ee
For a general 4-fold Fourier series as $h(\mathbf{x},t) = \sum_{n,m} a_{n,m}(t) e^{iq(nx+my)}$, this is nothing but
\be
\frac{1}{2} (n^4 + m^4) \ge n^2m^2
\ee
alias $(n^2-m^2)^2 \ge 0$, thus verifying the inequality.},
from Eq.~(\ref{autov.quad2}): both coefficients are evidently negative\footnote{The case $c_{11}=c_{12}$ would produce a one-dimensional dynamics, thus we will not take it into consideration.}, and they have the same $q$-dependence. Using (\ref{dimensional}) the coarsening exponent can be easily extracted:
\be
L \sim t^{1/3}.
\label{n1_3}
\ee

For models without slope selection, the current $\mathbf{j}$ has no zeros. A prototype of this kind of currents is asymptotically
represented by $\mathbf{j}(\mathbf{m})\simeq 1/|\mathbf{m}|^{\alpha}$, $\alpha>1$. Exploiting Eq.~(\ref{corr_tot}) and coupling it with the asympotic expression of the current, we obtain
\be \label{no_slope_selection}
-\nabla^2 \mathbf{m} \simeq 1/|\mathbf{m}|^{\alpha}.
\ee
We now switch to polar coordinates and make the assumption that mound profile changes only along one direction but remains constant along the perpendicular one, so that Eq.(\ref{no_slope_selection}) can be mapped
onto a one-dimensional equation: $m'' + (1/r) m'+ 1/m^{\alpha}$, neglecting the angular dependence for $\mathbf{m}$. Plugging in it a solution of the form $m \simeq A r^{\gamma}$, we find $\gamma=2/(\alpha+1)$ and
finally, calculating again 
the scalar products in Eqs. (\ref{autov.quad1}) and (\ref{autov.quad2}) for the
square symmetry case, we get
\begin{equation}
L \sim t^{1/4}
\label{n1_4}
\end{equation}
for any value of $\alpha$. 

We can gain further insight  from dimensional considerations: 
if $m \approx L^{2/\alpha+1}$, then $m \approx t^{\beta}=t^{2n/\alpha+1}$,
therefore giving $\beta=1/(2(\alpha+1))$.
We have found
that all the other symmetries hereby mentioned produce exactly the same exponents, $n$ and $\beta$, pointing to the existence of universality classes.
Our results are summarized in Table II.

It is worth comparing our results for $n$ with the corresponding values for the one-dimensional growth models. 
The models with constantly increasing slope yield $n=1/4$ \cite{1d,Politi_Torcini}, as we have also found in two dimensions.
The $1d$ model with faceting, instead, is known to produce a logarithmic coarsening in the absence of noise \cite{1d,Langer} and
$n=1/3$ when noise is present \cite{banzai,*banzai2}.
We conclude, with the caveat of noise and in analogy with the case of phase separation processes (see Sec.VII.B for a thorough discussion),
that the dimension of physical space, $d$, 
seems \footnote{We do not make a stronger claim because we limit the comparison to $d=1$ and $d=2$.} to be irrelevant for our class of growth models, Eq.~(\ref{eq.crescita}).

\begin{table}
$
\begin{array}{>{\displaystyle}c|>{\displaystyle}c|>{\displaystyle}c}
\mbox{current~} \bj \mbox{~producing} & ~~n~~ & ~~~\beta\\[1em]\hline
& & \\
\mbox{faceting} & \frac{1}{3} & ~~0\\[1em]\hline
& & \\
\mbox{increasing slope} & \frac{1}{4} & ~~\frac{1}{2(1+\alpha)}\\
\end{array}
$
\caption{The value of the coarsening exponent $n$ and of the exponent $\beta$ ($|\bm(t)| \approx t^\beta$ for large $t$,
for the two universality classes resulting from our study. 
The exponent $\alpha$ appears in the relation
$\bj(\bm) \approx 1/|\bm|^\alpha$ for large $\bm$. }
\end{table}

\section{Discussion}
\label{sec_discussion}

\subsection{The context}
\label{Sec_disc}

The field of crystal growth processes by a vapour phase has been very active in the past twenty-five years,
involving experiments, simulations, and analytic studies. General references of special interest for the
present article are three review papers~\cite{Politi_review,Misbah_review,Golubovic_review} and one book~\cite{Michely_Krug}.

Rigorous results for model equations included in the class (\ref{eq.crescita}) studied here are very rare.
We are aware of two exact inequalities, concerning isotropic currents and which are in agreement with our universality classes:
Kohn\&Yan~\cite{Kohn_Yan} studied the faceted case, finding $n\le 1/3$;
Bo\&Li~\cite{BoLi} studied the increasing slope case, finding $n\le 1/4$. 
Another worth mentioning paper is the study by Watson\&Norris~\cite{Watson_Norris} of the faceted case
with a three-fold symmetric current: authors find $n=1/3$, also in agreement with our findings.

Less rigorous results are often based on approximate evaluations of the different terms appearing in the
equation and on how such terms depend on scale length $L$. A significant amount of effort has been devoted to such an approach by Golubovi\'c and
collaborators~\cite{Golubovic_review}. While in the ``no faceted" case~\cite{Golubovic1997}
results are in agreement with ours, the faceted case is controversial. More precisely, in Ref.~\cite{Moldovan_Golubovic} the
six-fold symmetric case gives $n=1/3$, while in Ref.~\cite{Levandovsky_Golubovic} the four-fold case may give
either $n=1/3$ or $n=1/4$, depending on the details of the current.
Siegert too, already a few years before, had claimed to find a slower coarsening,
$n=1/4$, when integrating numerically a (\ref{eq.crescita})-like equation with square symmetry and faceted
morphology ~\cite{Siegert98}. 

The peculiar result $n=1/4$ for square symmetry would be related, according to the above mentioned authors, to the existence of two
different types of domain walls, pyramid edges and roof tops: in the former case only one component of the slope
changes, while in the latter case both components change. Roof tops, which are not present in a regular, periodic
square lattice, act as dislocations and would play a major role in their simulations, slowing down coarsening.
However, they also claim that the system would be frozen in the absence of roof tops, because the square pattern
would be metastable. The last statement is in contrast to our findings, according to which the square pattern is expected to be linearly
unstable. 
We think that the square case would require some more analysis to gain more insights on the role of defects,
as well as on the roles of the specific form of the current, of the initial conditions and of the simulation
time. 
For the sake of completeness we also make our reader aware that the same considerations just expressed for the square case might be extended to the less studied rectangular case, since a different coarsening exponent can be found in the literature ($n=2/7$) \cite{Golubovic_review}. Again, also for this symmetry, coarsening is bound to the presence of dislocations and the specific form of the current might be relevant (e.g., its derivability from a potential).

While comparing with previous studies of the same class of equations is straightforward,
comparing with Kinetic Monte Carlo simulations and with experiments becomes difficult and dangerous,
mainly for two reasons: (i)~Is the system (the simulated system or the real system) described
by a (\ref{eq.crescita})-like equation or different equations are more appropriate?
(ii)~Does the simulation or the experiment attain large enough times to probe the asymptotic regime?
Honestly, the variety of results of simulations/experiments and the above two questions prevent from giving a clear
picture of such results and from connecting them to specific model equations.
As for experiments, we refer the reader to Table 2 or Ref.~\cite{Politi_review} and to Table 4.2 of 
Ref.~\cite{Michely_Krug}. We close this Section by giving a few more details on point (i), here above.

Equation~(\ref{eq.crescita}) is a class of general models because $\bj(\bm)$ has the only requirement to be linear
at small slopes. Such equation cannot cover any possible model of growth by vapor deposition. For example,
some studies have suggested a higher order linear term, which would produce slower coarsening~\cite{Golubovic1997}
(if two linear terms of different order are present, crossover effects are expected, depending on their relative
strength). A second remark is related to the up-down symmetry of the emerging morphology. 
Eq.~(\ref{eq.crescita}) is invariant under the transformation
$h\to -h$ but such a symmetry is weak or absent in simulations and experiments. Therefore, symmetry breaking terms have
been introduced, for example in Ref.~\cite{Stroscio}. While in $d=1$ such a term
seems to be irrelevant~\cite{PPkinks}, the question of its relevance in $d=2$ is still open.
A final comment concerns the form of $\bj(\bm)$. Whatever its symmetry, we have assumed it is analytical at
$\bm=0$. However, some results~\cite{Politi_Krug} suggest that step-edge diffusion might produce a current which
is singular at vanishing slope.

\subsection{Crystal growth vs phase separation}

In the previous Section we have discussed how our results compare with other studies on
crystal growth. Here we rather focus on differences and similarities with phase separation processes.

First of all,
the irrelevance of space dimensionality we have found to be valid for our growth model Eq.(\ref{eq.crescita}) is also a well-known feature of
phase-separation processes, as long as $T_c > 0$. In such context,
for a scalar order parameter the coarsening exponents are $n = 1/2$ for a nonconserved
order parameter and $n = 1/3$ for a conserved
one \cite{Bray,*Bray_Rutenberg,1d}. 
However, in $d = 1$, such figures are found when
noise is present \cite{banzai,*banzai2}, otherwise a slow logarithmic coarsening
appears \cite{Langer}. The one dimensional growth model
with faceting is equivalent to a conserved phase separation
process (so called model B of dynamics or Cahn-
Hilliard equation). Therefore, it is not surprising that
it gives logarithmic coarsening without noise
and $n = 1/3$ in the presence of noise. 
However, the analogy with phase separation processes cannot be pushed further, because our growth model has, in $2d$, peculiar proprieties. 

Even if 
our multiscale approach is applicable, in principle, to any nonequilibrium current
$\bj(\bm)$, we have considered here the case of symmetric Jacobian matrix ${\cal J}$,
which means $\partial j_i/\partial m_\ell=\partial j_\ell/\partial m_i$.
This condition is satisfied by {\it all} crystal growth equations
we are aware of and it is equivalent to saying that dynamics can be cast into a variational formulation. 
In fact, Eq.~(\ref{eq.crescita}) can be rewritten as:
\be
\label{eq_potential}
\frac{\partial h}{\partial t} = \nabla \cdot \frac{\delta {\cal F}}{\delta \bm} ,
\ee
if and only if ${\cal J}$ is symmetric,
where
\be
{\cal F} = \int d\bx \left\{ \frac{1}{2}\left[ (\nabla m_x)^2 + (\nabla m_y)^2\right] +
V(\bm)\right\},
\ee
so that
\be
\begin{split}
\frac{d {\cal F}}{dt} &= \int d\bx \frac{\delta {\cal F}}{\delta m_i}
\frac{\partial m_i}{\partial t} =
\int d\bx \frac{\delta {\cal F}}{\delta m_i} \frac{\partial}{\partial x_i}
\left(\nabla \cdot \frac{\delta {\cal F}}{\delta \bm}\right)\\
&= -\int d\bx \left( \nabla \cdot \frac{\delta {\cal F}}{\delta \bm} \right)^2 \le 0 .
\end{split}
\ee
Taking the gradient of both sides of Eq.~(\ref{eq_potential})
we obtain
\be
\label{eq_m_growth}
\frac{\partial \bm}{\partial t} = \nabla\left(
\nabla \cdot \frac{\delta {\cal F}}{\delta \bm} \right) ,
\ee
which is reminiscent of the B-dynamics for a conserved, vector order parameter:
\be
\label{eq_B_dynamics}
\frac{\partial \bm}{\partial t} = \nabla^2 \left(
\frac{\delta {\cal F}}{\delta \bm} \right) .
\ee
The very first remark is that we can evoke some similarity with a
phase separation process only in the case of faceting.
In fact, phase separation requires that $V(\bm)$ has minima
for finite $\bm$ and these minima correspond to the {\it magic slopes}
for which $\bj$ vanishes and which keep constant in time during
the coarsening process.
The second remark is that Eqs. (\ref{eq_m_growth}) and (\ref{eq_B_dynamics}) 
are different (see what $\nabla$ applies to) and the
order parameter $\bm$ is also different: in Eq.~(\ref{eq_m_growth})
$\bm =\nabla h$, which means $\nabla\times\bm =0$. 
In simple terms, $\bm =\nabla h$ implies that domain walls must be
straight lines, because they result from the intersection of two 
planes (regions of constant $\bm$), 
while the shape of domain walls in standard phase separation processes
has no such constraint.

Because of these differences between 
Eqs. (\ref{eq_m_growth}) and (\ref{eq_B_dynamics}),
it should not appear surprising they give different coarsening exponents.
While the crystal growth equation~(\ref{eq_m_growth}) gives
$n=\frac{1}{3}$ irrespectively of the symmetry of $V(\bm)$,
for the phase separation process, Eq.~(\ref{eq_B_dynamics}),
we have $n=\frac{1}{4}$ if $V(\bm)$ has a continuous family of minima
(i.e. an infinite number of minima corresponding to a circularly symmetric current),
and $n=\frac{1}{3}$ if $V(\bm)$ has a finite number of minima (the current has rotational symmetry under a specific angle),
because such case corresponds to a scalar order 
parameter~\cite{BrayPRL,SiegertPhysica}.

In conclusion, we have proposed a classification of important unstable crystal growth dynamics in terms of universality classes, detecting what features are relevant (faceting or not) and what features are irrelevant (symmetry of the pattern, the symmetry of the surface mass current and the space dimensionality). Therefore, we have shown that Eq.~(3) has distinct properties and critical exponents, conferring to unstable crystal growth a place as novel nonequilibrium paradigm.

\begin{acknowledgments}
CM thanks CNR for a Short-Term Mobility award from their International Exchange
Program and CNES for financial support.
\end{acknowledgments}

\appendix

\section{Growth equation perturbative expansions}
\label{operators}
Using (\ref{time}) and (\ref{space}), we can express different operators of our growth model equation (\ref{eq.crescita}).
For the Laplacian we have:
\begin{equation} \label{Laplac}
\nabla^2 = \nabla_0^2 + \varepsilon [\nabla_0 \nabla_\mathbf{X} + \nabla_\mathbf{X} \nabla_0]
\equiv \nabla_0^2 + \varepsilon \nabla_1^2 .
\end{equation}
The expansion of the  current takes the form:
\begin{equation}
\begin{split}
\mathbf{j}(\nabla h) =&
\mathbf{j}(\nabla_0 (\th_0 + \varepsilon \th_1) + \varepsilon \nabla_\mathbf{X} (\th_0 + \varepsilon \th_1))=
\\
=& \mathbf{j}(\nabla_0 \th_0) + \varepsilon \mathcal{J} (\nabla_0 \th_1 + \nabla_\mathbf{X} \th_0) \nonumber ,
\end{split}
\end{equation}
with $\mathcal{J}$ being the Jacobian matrix of $\mathbf{j}$, and:
\begin{equation}
\nabla^2 h=(\nabla_0^2 + \varepsilon \nabla_1^2) (\th_0 + \varepsilon \th_1) = \nabla_0^2 \th_0 + \varepsilon (\nabla_1^2 \th_0 + \nabla_0^2 \th_1) \nonumber
\end{equation}
with its gradient:
\begin{equation}
\nabla (\nabla^2 h) = \nabla_0 (\nabla_0^2 \th_0) + \varepsilon [\nabla_\mathbf{X} (\nabla_0^2 \th_0) + \nabla_0 (\nabla_1^2 \th_0 + \nabla_0^2 \th_1)] \nonumber.
\end{equation}
Thus, our model equation (\ref{eq.crescita}) becomes at first order:
\be
\begin{split}
\varepsilon & [(\partial_T \psi_1) \partial_{\varphi_1} \th_0 + (\partial_T \psi_2) \partial_{\varphi_2} \th_0]
= \\
= & -(\nabla_0+\varepsilon \nabla_{\mathbf{X}}) \cdot \{  \mathbf{j}(\nabla_0 \th_0) +\varepsilon \mathcal{J} (\nabla_0 \th_1 + \nabla_\mathbf{X} \th_0) \\
+ & \nabla_0 (\nabla_0^2 \th_0) + \varepsilon [\nabla_\mathbf{X} (\nabla_0^2 \th_0) + \nabla_0 (\nabla_1^2 \th_0 + \nabla_0^2 \th_1)] \} .
\end{split}
\ee

\section{Expression for the adjoint $\mathcal{L}^{\dag}$}
\label{adjoint}
From the first order contribution (\ref{linear}) of our growth equation we define the linear operator $\mathcal{L}$:
\begin{equation}
\mathcal{L}[u]=-\nabla_0 \cdot \mathcal{J}\nabla_0 u - \nabla_0^2 (\nabla_0^2 u)
\end{equation}
In order to determine the adjoint operator we split the above linear operator into two parts
\bea
\langle \mathcal{L^{\dag}}_1 v,u \rangle &=& -\langle v,\nabla_0 \cdot \mathcal{J}\nabla_0 u \rangle \\  \langle \mathcal{L^{\dag}}_2 v,u \rangle &=& -\langle v, \nabla_0^2 (\nabla_0^2 u) \rangle .
\eea
and apply the definition $\langle \mathcal{L^{\dag}}v,u \rangle = \langle v,\mathcal{L} u \rangle$ for each term separately.
Integration by part in the second term yields:
\begin{equation}
\langle \mathcal{L^{\dag}}_2 v,u \rangle = -\langle v, \nabla_0^2 (\nabla_0^2 u) \rangle = -\langle \nabla_0^2 v, \nabla_0^2 u \rangle = -\langle \nabla_0^4 v, u \rangle , \nonumber
\end{equation}
(integrals over the surface vanish because of periodicity of the integrand over the interval $[0,2\pi]$). Thus $\mathcal{L}_2$ is self-adjoint.
Regarding the first term
 ${\cal L}_1$ it is more convenient to use explicitly the integral notation for the scalar product and express it in terms of  components:
\be
\begin{split}
\langle \mathcal{L^{\dag}}_1 v,u \rangle &= -\langle v,\nabla_0 \cdot \mathcal{J}\nabla_0 u \rangle \equiv \\
& \equiv -\frac{1}{(2\pi)^2} \int\int v^* [(\nabla_0)_i \mathcal{J}_{ij} (\nabla_0)_j u ]=\\
&=+ \frac{1}{(2\pi)^2} \int\int [(\nabla_0)_i v^*] \mathcal{J}_{ij} [(\nabla_0)_j u]= \\
&= +\frac{1}{(2\pi)^2} \int\int \mathcal{J}_{ij} [(\nabla_0)_i v^*] [(\nabla_0)_j u]= \\
&= - \frac{1}{(2\pi)^2} \int\int [(\nabla_0)_j \mathcal{J}_{ij} (\nabla_0)_i v^*] u ,
\end{split}
\ee
i.e.
\begin{equation}
\mathcal{L^{\dag}}_1 \!=\! -(\nabla_0)_i \mathcal{J}_{ji} (\nabla_0)_j \!=\! -(\nabla_0)_i \mathcal{J}_{ij}^T (\nabla_0)_j \! \equiv \! -\nabla_0 \cdot \mathcal{J}^T \nabla_0 . \nonumber
\end{equation}
Therefore the total $\mathcal{L}$ operator is \textit{self-adjoint} if and only if the Jacobian matrix of the current $\mathbf{j}$ is symmetric.

\section{Solvability conditions}
\label{c_abg}
Let us write down explicitly the solvability equations (\ref{cond.risolv}):
\bea
g &\equiv& (\partial_T \psi_1) h_1 + (\partial_T \psi_2) h_2 + \nabla_0 \mathcal{J}\nabla_{\mathbf{X}} h_0 + \nabla_0 \nabla_{\mathbf{X}} \nabla_0^2 h_0 \nonumber \\
&+& \nabla_0^2 (\nabla_0 \nabla_{\mathbf{X}} +\nabla_{\mathbf{X}} \nabla_0) h_0
\eea
and explicit out the calculation on terms entering the right hand side of the above equation. Using $\partial/\partial X_{\gamma}=(\partial q_{\alpha\beta}/\partial X_{\gamma})
\partial/\partial q_{\alpha\beta}=(\partial \psi_{\alpha}/\partial X_{\gamma} \partial X_{\beta})
\partial/\partial q_{\alpha\beta} $, we can write:
\begin{align}
&\nabla_0 \mathcal{J}\nabla_{\mathbf{X}} h_0 =
\frac{\partial \psi_{\alpha}}{\partial X_{\gamma} \partial X_{\beta}} q_{il} \partial i \left( \mathcal{J}_{l\beta} \frac{\partial h_0}{\partial q_{\gamma \alpha}} \right) \nonumber , \\
&\nabla_0^2 \nabla_0 \nabla_{\mathbf{X}} h_0 = 
\frac{\partial \psi_{\alpha}}{\partial X_{\gamma} \partial X_{\beta}} \nabla_0^2 q_{i\gamma}
\frac{\partial h_i}{\partial q_{\alpha \beta}} , \nonumber \\
&\nabla_0^2 \nabla_{\mathbf{X}} \nabla_0 h_0 = 
\frac{\partial \psi_{\alpha}}{\partial X_{\gamma} \partial X_{\beta}}
\left( \nabla_0^2 \delta_{i\alpha} \delta_{\gamma \beta} h_i + \nabla_0^2 q_{i\gamma} \frac{\partial h_i}{\partial q_{\alpha \beta}} \right),  \nonumber \\
&\nabla_0 \nabla_{\mathbf{X}} \nabla_0^2 h_0 =
\frac{\partial \psi_{\alpha}}{\partial X_{\gamma} \partial X_{\beta}} 2 q_{i\gamma}  \delta_{j\alpha} \delta_{l\beta} q_{nl} h_{ijn} \nonumber \\
& \qquad \qquad \quad + \frac{\partial \psi_{\alpha}}{\partial X_{\gamma} \partial X_{\beta}} q_{i\gamma} \nabla_0^2 \frac{\partial h_i}{\partial q_{\alpha \beta}}  \nonumber ,
\end{align}
where $\delta$'s arise from derivatives: $\delta_{sl} \delta_{pt}= \partial q_{sp}/\partial q_{lt}$.

\section{Diffusion coefficients for the hexagonal symmetry}
\label{hex.diff}
Our aim here is to prove that $D_{11}^{11} \equiv D_{22}^{22} \equiv D_{11}$ for hexagonal symmetry.
We show detailed calculations in this specific case only, that can serve as a guide for other symmetries.
We use definitions (\ref{convenz.coeff}) in order  to write the two coefficients $D_{\beta\gamma}^{i\alpha}$. First  we write down  some of the $c^{\alpha}_{\beta \gamma}$'s that enter the diffusion coefficients:
\bea
&& -c^1_{11} = q[(\partial_1+\partial_2)(\mathcal{J}_{11}\partial_q h_0)+2\sqrt{3}
(\partial_1 - \partial_2)(\mathcal{J}_{21}\partial_q h_0)] \nonumber \\
&&+ \frac{q^2}{2} (h_{111}+ 2 h_{112}+h_{122}) +3 \nabla_0^2 q(\partial_1+\partial_2) \partial_q h_0+ \nabla_0^2 h_1 , \nonumber
\eea
and:
\bea
&& -c^2_{22} =
-q[\frac{1}{\sqrt{3}}(\partial_1+\partial_2)(\mathcal{J}_{12} \partial_q h_0)+
(\partial_1 - \partial_2) (\mathcal{J}_{22} \partial_q h_0)] \nonumber \\
&&+ \frac{3}{2}q^2 (h_{112}+ h_{222}) -3 \nabla_0^2 q(\partial_1-\partial_2) \frac{\partial h_0}{\partial q}+ \nabla_0^2 h_2 . \nonumber
\eea
For the Jacobian matrix components we use  condition (\ref{corr_tot}), that amounts to:
\begin{equation}
\mathbf{j}(\nabla_0 h_0) = - \nabla_0(\nabla_0^2 h_0) ,
\end{equation}
allowing to obtain:
\begin{widetext} \label{Jtrian}
\begin{gather}
\mathcal{J}_{11} = \frac{q^2}{2} \left[ \frac{(h_{12}-h_{22})(h_{1111}+h_{1222}) - (h_{11}-h_{12})(h_{1112}+h_{2222})}
{h_{11} h_{22} - h_{12}^2} \right] , \nonumber \\
\mathcal{J}_{12} = \frac{q^2}{2\sqrt{3}} \left[ \frac{(h_{11}+h_{12})(h_{1112}+h_{2222}) - (h_{12}+h_{22})(h_{1111}+h_{1222})}
{h_{11} h_{22} - h_{12}^2} \right] ,\nonumber \\
\mathcal{J}_{21} = \frac{\sqrt{3}q^2}{2} \left[ \frac{(h_{12}-h_{11})(h_{1112}-2h_{1122}+2h_{1222}-h_{2222}) + (h_{12}-h_{22})(h_{1111}+2h_{1122}-h_{1222}-2h_{1112})}{h_{11} h_{22} - h_{12}^2} \right] , \nonumber \\
\mathcal{J}_{22} = \frac{q^2}{2} \left[ \frac{(h_{11}+h_{12})(h_{1112}-2h_{1122}+2h_{1222}-h_{2222}) + (h_{22}+h_{12})(h_{1222}+2h_{1112}-h_{1111}-2h_{1122})}{h_{11} h_{22} - h_{12}^2} \right] . \nonumber
\end{gather}
Moreover, $\langle h_1^2\rangle  = \langle h_2^2\rangle  \equiv \langle h_{\varphi}^2\rangle $.
We can now write
\begin{gather} \label{D11.esag}
D_{11}^{11}=\frac{1}{\langle h_1h_2\rangle } [q^3 \partial_q \langle h_{11}^2 \rangle + q^2\langle h_{11}^2\rangle + \frac{q^2}{2} \langle h_{11} h_{22} \rangle] , \\
D_{22}^{22}=\frac{1}{\langle h_1h_2\rangle }\left \{ 4 q^3 \left[ \frac{1}{2}\partial_q \langle h_{11}^2 \rangle -\frac{3}{2} \partial_q \langle h_{11} h_{12}\rangle + \partial_q \langle h_{11} h_{22}\rangle \right] + q^2 \left[ \frac{7}{2} \langle h_{11}h_{22}\rangle - \frac{1}{2} \langle h_{11}^2\rangle \right] \right\}.
\end{gather}
\end{widetext}
Invoking invariances under $\pi/3$, we find that specific and non-trivial relations are verified among scalar products:
\begin{equation} \label{inattese}
\langle h_{11}^2\rangle =2\langle h_{11}h_{12}\rangle , \quad \langle h_{11}h_{12}\rangle =\langle h_{11}h_{22}\rangle.
\end{equation}
By means of the first relation of (\ref{inattese}), we finally get:
\begin{equation}
D_{11}^{11} - D_{22}^{22} = \frac{1}{h_1h_2}(3q^2+4q^3\partial_q)[\langle h_{11}h_{12}\rangle -\langle h_{11}h_{22}\rangle ] \nonumber
\end{equation}
that vanishes thanks to the second relation (\ref{inattese}), therefore $D_{11}^{11}=D_{22}^{22}$.

\section{$\mathbf{q}$-component for various symmetries}
\label{q_vectors}
We declare the $\mathbf{q}$-vectors used in the treated cases:
\begin{itemize}
\item {\it rectangular}: \quad $\mathbf{q}_1=q(1,0)$, $\mathbf{q}_2=pq(0,1)$
\item {\it square}: \quad $\mathbf{q}_1=q(1,0)$, $\mathbf{q}_2=q(0,1)$
\item {\it hexagonal}: \quad $\mathbf{q}_1=q/2(1,\sqrt{3})$, $\mathbf{q}_2=q/2(1,-\sqrt{3})$
\item {\it triangular}: \quad $\mathbf{q}_1=q/2(1,\sqrt{3})$, $\mathbf{q}_2=q/2(1,-\sqrt{3})$
\end{itemize}

\section{Phase equation eigenvalue spectrum for the rectangular case}
\label{rectangular}
The eigenvalue sign for the rectangular case spectrum in Eq.~(\ref{Omega_rec}) is not easy to recognize because of the $\theta$-dependence. Thanks to continuity in $\theta$, however, we limit the analysis
only to the extremal values, namely the $\theta$ directions such that $\partial \Omega_{1,2}/\partial \theta =0$.
Let's use the notation $s \equiv \sin^2{\theta}$ and let's define $\Omega_{1,2}(K,s)=G(s)K^2/2$ with
\be
G(s)=-\big[(D^{11}_{11}+D^{22}_{11}) + As \big] \pm \sqrt{D+Bs+Bs^2} ,
\ee
where
\bea \label{alphabet}
A &:=& D^{11}_{22} + D^{22}_{22} - D^{11}_{11} - D^{22}_{11}  , \nonumber \\
B &:=& 2 (D^{11}_{11} \!-\! D^{22}_{11}) (D^{11}_{22} - D^{22}_{22} - D^{11}_{11} + D^{22}_{11})\!+\! 4 D^{12}_{12} D^{21}_{12}  , \nonumber \\
C &:=& ( D^{11}_{22} - D^{22}_{22} - D^{11}_{11} + D^{22}_{11} )^2 - 4 D^{12}_{12} D^{21}_{12}  ,  \\
D &:=& ( D^{11}_{11} - D^{22}_{11} )^2 \nonumber .
\eea
We now write an analogous extremal condition for Eq.~(\ref{Omega_rec}):
\be \label{rect.extreme}
\frac{\partial G(s)}{\partial s} \frac{\partial s}{\partial \theta} =0.
\ee
The annihilation of the second factor reveals that $\theta=0+n\pi$ and $\theta=\pi/4 + n\pi$ are extremal directions, the annihilation of the first factor gives further solutions:
\bea \label{sin.quadri}
&s& = \frac{1}{2C (A^2-C)} \big\{-B(A^2-C)  \\
&\pm& \sqrt{B^2(A^2-C)^2- 4C(A^2-C)(A^2 D -B^2/4)} \big\}. \nonumber
\eea
A check of its sign and value, if not feasable in general terms, is possible in the limit of small amplitude for the growing pattern, as we are going to proof.
Given the symmetries indicated in Table I for the rectangular patern and following the same steps indicated in Section V for the square symmetry case,
a rectangular Fourier series is, at first order ($n,m=\pm 1$):
\bea \label{rect.Fourier}
h(\mathbf{x},t) &\simeq& a_1(t) (e^{iq(x+py)} + e^{iq(x-py)} + c.c.) = \nonumber \\
&=& 4 a(t) \cos\varphi_1 \cos\varphi_2
\eea
where $a_{1,1}=a_{1,-1}=a_{-1,1}=a_{-1,-1} \equiv a$. From the second expression in Eq.~(\ref{rect.Fourier}) is straightforward to calculate the scalar products appearing in the six diffusion coefficients,
obtaining the following approximated forms:
\bea
D_{11}^{11} &=& \frac{1}{\langle h_1^2\rangle } [ \partial_q (q^3 \langle h_{11}^2 \rangle)  + q^3 p^2 \partial_q \langle h_{12}^2\rangle + q^2 p^2 \langle h_{12}^2\rangle ] \nonumber \\
&\simeq& (3+p^2)q^2 + (1+p^2) q^3 \partial_q (a^2)/a^2, \nonumber \\
D_{22}^{22} &=& \frac{1}{\langle h_2^2\rangle } [p^2 \partial_q (q^3 \langle h_{22}^2\rangle) + q^3 \partial_q \langle h_{12}^2 \rangle  + q^2 \langle h_{12}^2\rangle ] \nonumber \\
&\simeq& (3p^2 +1)q^2 + (1+p^2) q^3 \partial_q (a^2)/a^2, \nonumber \\
D_{22}^{11} &=& \frac{q^2}{\langle h_1^2\rangle } [\langle h_{11}^2\rangle  + 3 p^2 \langle h_{12}^2\rangle ] \simeq (3+p^2)q^2, \nonumber \\
D_{11}^{22} &=& \frac{q^2}{\langle h_2^2\rangle } [3 \langle h_{12}^2\rangle  + p^2 \langle h_{22}^2\rangle ] \simeq (1+3p^2)q^2 ,\nonumber \\
D_{12}^{12} &=& \frac{2}{\langle h_1^2\rangle } [p q^3 \langle h_{11} \partial_q h_{12} \rangle  + p^3 q^3 \langle h_{22} \partial_q h_{12} \rangle + pq^2 \langle h_{12}^2 \rangle ] \nonumber \\
&\simeq& 2pq^2,  \nonumber \\
D_{21}^{12} &=& \frac{2}{\langle h_2^2\rangle } [q^3/p \langle h_{11} \partial_q h_{12} \rangle  + p q^3 \langle h_{22} \partial_q h_{12} \rangle + 2 pq^2 \langle h_{12}^2 \rangle ] \nonumber \\
&\simeq& 4pq^2. \nonumber
\eea
Let's indicate, for the sake of brevity, $m \equiv q^3 \partial_q (a^2)/a^2$ and let's rewrite Eqs.(\ref{alphabet}):
\bea
A &:=& 4(p^2-1) q^2 , \nonumber \\
B &:=& -4 (1+p^2)^2 m^2 + 64p^2 q^2 , \nonumber \\
C &:=& 4(1+p^2)^2 m^2 -32p^2q^2  ,  \\
D &:=& (1+p^2)^2 m^2 \nonumber .
\eea
In the weakly nonlinear regime Eqs.~(\ref{sin.quadri}) become:
\be
s=\frac{1}{2} \pm 2 \sqrt{2} \frac{q^4}{m^2} \frac{p^2-1}{(p^2+1)^2} p .
\ee
We note that for $p=1$ (square symmetry case), these two directions coincide in $\theta=\pi/4$.
The corresponding eigenvalues can be now written in their approximated expressions; for completeness we list all the $\Omega_{1,2}(K)$ found for the rectangular case spectrum Eq.~(\ref{Omega_rec}):
\bea
&\Omega&_1^0 (K,p,q) \simeq -(3+p^2) q^2 K^2 , \nonumber \\
&\Omega&_2^0 (K,p,q) \simeq -(1+p^2) m K^2 ,  \nonumber \\
&\Omega&_1^{\pi/2} (K,p,q) \simeq -(1+3p^2) q^2 K^2 ,  \nonumber \\
&\Omega&_2^{\pi/2} (K,p,q) \simeq -(1+p^2) m K^2 ,  \nonumber \\
&\Omega&_1^{s+} \! (K,p,q)\! =\! \Omega_2^{s+} (K,p,q) \!=\! \Omega_1^{s-} (K,p,q) \!=\! \Omega_2^{s-} (K,p,q) \!=\! \nonumber \\
&\simeq& -(1+p^2) m \frac{K^2}{2} .  \nonumber
\eea
This weakly nonlinear analysis allows finally to recognize the positivity of $\Omega_1^0 (K,p,q)$ and $\Omega_1^{\pi/2} (K,p,q)$, already manifest in Eqs.~(\ref{posit_rect}),
and to state how the sign of all the other eigenvalues
remains undoubtedly determined by the behaviour of the stationary amplitude with respect to the stationary wavelength $\lambda=2\pi/q$. In Section V a further step, explicitly shown for
the square symmetry case but still valid also for the rectangular symmetry,
allows also to proof that the behaviour of $\partial_q (a)$
is related to the specific form of the current $\mathbf{j}$.

\bibliography{biblio1}

\comment{

}

\end{document}